\documentclass[11pt]{article}
\usepackage[preprint]{acl}

\usepackage{times}
\usepackage{latexsym}
\usepackage{float}
\usepackage{ragged2e}
\usepackage{dblfloatfix}   
\usepackage{placeins}  
\usepackage{needspace}

\usepackage[T1]{fontenc}

\usepackage[utf8]{inputenc}

\usepackage{microtype}

\usepackage{inconsolata}

\usepackage{graphicx}

\title{Towards an Automated Framework to Audit Youth Safety on TikTok}

\author{
 \textbf{Linda Xue\textsuperscript{1}},
 \textbf{Francesco Corso\textsuperscript{2}},
 \textbf{Nicolo' Fontana\textsuperscript{2}},
 \textbf{Geng Liu\textsuperscript{2}},
\\
 \textbf{Stefano Ceri\textsuperscript{2}},
 \textbf{Francesco Pierri\textsuperscript{2}}
\\
\\
 \textsuperscript{1}Massachusetts Institute of Technology,
 \textsuperscript{2}Politecnico di Milano
\\
 \small{
   \textbf{Correspondence:} \href{mailto:email@domain}{lexue28@mit.edu}, \href{mailto:email@domain}{francesco.corso@polimi.it}, \href{mailto:email@domain}{nicolo.fontana@polimi.it}, \href{mailto:email@domain}{geng.liu@polimi.it}, 
    }
    \\
    \small{   
       \href{mailto:email@domain}{stefano.ceri@polimi.it}, \href{mailto:email@domain}{francesco.pierri@polimi.it}  
    }
}

\begin{document}

\maketitle

\begin{abstract}
This paper investigates the effectiveness of TikTok's enforcement mechanisms for limiting the exposure of harmful content to youth accounts. We collect over 7000 videos, classify them as harmful vs not-harmful, and then simulate interactions using age-specific sockpuppet accounts through both passive and active engagement strategies. We also evaluate the performance of large language (LLMs) and vision-language models (VLMs) in detecting harmful content, identifying key challenges in precision and scalability. 

Preliminary results show minimal differences in content exposure between adult and youth accounts, raising concerns about the platform's age-based moderation. These findings suggest that the platform needs to strengthen youth safety measures and improve transparency in content moderation.
\end{abstract}

\section{Introduction}

TikTok is a short-form video platform that has rapidly emerged as one of the world's most influential social media services. With over 1.6 billion monthly active users\footnote{\url{https://www.businessofapps.com/data/tik-tok-statistics}, accessed on August 5, 2025} and millions of videos uploaded daily~\citep{corso2024we}, it now plays a central role in the global digital media landscape. 

Children and adolescents increasingly rely on TikTok for both entertainment and everyday information~\citep{violot2024shorts, liu2024usage, ge2021effect}. Although TikTok enforces community guidelines through content removal and age-based restrictions \footnote{\url{https://www.tiktok.com/community-guidelines}, accessed on August 8, 2025}, concerns remain about the effectiveness of these moderation mechanisms in shielding young users from harmful content. These concerns are amplified by the introduction of the European Digital Services Act \footnote{\url{https://commission.europa.eu/strategy-and-policy/priorities-2019-2024/europe-fit-digital-age/digital-services-act_en}}, which requires very large online platforms to assess and mitigate systemic risks---particularly those related to the protection of minors and the spread of harmful content.




This paper reports preliminary findings from an ongoing work aiming to build an automated pipeline for auditing TikTok's safety enforcement mechanisms by systematically measuring the exposure to harmful content among youth and adult users, investigating different modes of interaction. Specifically, we explore two main research questions: 
\begin{itemize}
\item \textbf{RQ1:} What level of harmful content will an adult versus a minor be exposed to on their FYF? 
\item \textbf{RQ2:} Does actively searching for harm-adjacent keywords increase exposure to inappropriate content?
\end{itemize}
To this end, we created 10 Youth and 10 Adult sockpuppet accounts and simulated multiple sessions over several days, collecting over 7,000 videos across both ``For You Feed'' (FYF) scrolling and using active search with harm-adjacent keywords. 

In addition, as a precise detection of harmful content is an heavy task for humans, we also consider a third research question:
\begin{itemize}
\item \textbf{RQ3:} How effective are Large Language Models such as GPT-4o and VideoLLaMA3 at detecting harmful content?
\end{itemize}

To this end, we employed both text-only and video-based LLMs to estimate the presence of harmful content that violates TikTok's community guidelines and evaluated their performance. 


\section{Related Work}


TikTok's \textsc{For} \textsc{You} \textsc{Feed} (FYF) algorithm personalizes content based on user language, location, posting time, and interactions such as likes and follows~\citep{boeker2022personalization}. This personalization engine rapidly amplifies interest-aligned content—often within just 200 recommendations—thereby fostering echo chambers and limiting content diversity~\citep{baumann2025dynamics}.
 

These concerns are especially salient given TikTok's young user base—over 60\% of users were under 30 in 2021~\cite{iqbal2022tiktok}. A large-scale audit using more than 100 automated accounts found that watch time plays a central role in shaping recommendations, reinforcing problematic content loops through prolonged exposure~\cite{wsj2021tiktok}.

Recent studies further contextualize harm by examining user behavior. For instance, the median user consumes approximately 90 videos daily~\cite{zannettou2024}, while moderately addicted users average 7.86 minutes per session~\cite{yang2025studying}. An experimental audit comparing TikTok, YouTube, and Instagram showed that accounts registered as 13-year-olds encountered harmful content more frequently and rapidly than accounts of older users~\cite{eltaher2025protecting}.

\section{Methods}


\subsection{Data Collection}


To investigate how harmful content varies by user age and interaction mode, we created 20 TikTok accounts using the platform's web version: 10 accounts were set with an age of 13 (Youth) and 10 with an age slightly above 18 (Adult). All accounts were registered in Italy. These age values were chosen as they represent the boundary between TikTok's definition of ``youth'' (under 18) and adulthood\footnote{\url{https://support.tiktok.com/en/account-and-privacy/account-privacy-settings/privacy-and-safety-settings-for-users-under-age-18}}. 

Using a script we built \footnote{\url{https://anonymous.4open.science/r/tiktok-scraper-8424}} to scrape data from the TikTok website, we collected data over four consecutive days -- Thursday through Sunday -- to capture differences across both weekdays and weekends. For each account on each day, four browsing sessions were conducted, each containing 22 videos, totaling 88 videos per account per day. This approximates the average daily video exposure (89.9 videos) and simulates moderately engaged users, based on prior TikTok usage studies reporting ~27 minutes total watch time and ~7.86 minutes per session~\citep{yang2025studying}. In total, we collected over 7,000 videos across 20 accounts over the four-day period. Our dataset includes metadata for every collected video, such as the description text, hashtags, and engagement statistics (views, likes, comments, shares, etc.). For each video, we also retrieved the top 10 comments via HTTP requests to analyze user interactions.

We implemented two primary user interaction modes:

\textbf{Passive Scrolling:} Simulated natural browsing behavior by programmatically loading videos from the FYF, with randomized delays (10–20 seconds) between each request to mimic scrolling. No user input was provided beyond passive viewing, aligning with typical user consumption patterns.

\textbf{Active Searching:} For each harmful content category, we extracted three keywords based on TikTok's Youth Safety and Well-Being Guidelines. We used these keywords in the  search bar to retrieve videos potentially related to sensitive topics using one Adult and one Youth account. A complete list of keywords is provided in Appendix~\ref{sec:appendix}, Table~\ref{tab:keywords}.

Interestingly, sometimes keywords such as ``alcohol'' were censored in English for youth accounts, whereas its Italian equivalent ``Alcol'' still yielded search results though explicitly stated as age-inappropriate in the guidelines.
    


\subsection{Harmful Content Detection}
Our framework for identifying harmful content is grounded in TikTok's official Community Guidelines, particularly the sections related to youth safety. Categories such as sexual content, suicide/self-harm, and physical violence were prioritized. Closely related categories (e.g., youth and adult sexual abuse) were merged, while those that may have a less critical societal impact (e.g., animal abuse~\footnote{\url{https://www.healthdata.org/research-analysis/diseases-injuries-risks/factsheets-hierarchy}}) were excluded. See Table~\ref{tab:categories} in Appendix A for details.
We employed three methods for detecting harmful content:

\textbf{1) Textual Analysis:} We used the multilingual Detoxify\footnote{\url{https://github.com/unitaryai/detoxify}} model to evaluate the toxicity of the top 10 comments per video. Detoxify outputs scores in the range $[0, 1]$ to quantify the probability of a comment being toxic or not. We then fed all video descriptions to GPT-4o \footnote{\url{https://platform.openai.com/docs/models/gpt-4o}}, prompting the model to classify a video based on TikTok's harmful content guidelines.
 
\textbf{2) Visual Content Analysis:} We tested the performance of VideoLLaMA3 \footnote{\url{https://huggingface.co/DAMO-NLP-SG/VideoLLaMA3-7B}} by using it on a random selection of 100 videos. A custom prompt was used to assess visual and audio cues based on the same guideline framework as the GPT-based description classification.

\textbf{3) Manual Evaluation:} The same 100 videos was manually labeled by three native Italian speakers. Each reviewer independently categorized the content using our framework. Disagreements between two annotators were resolved by the third reviewer, ensuring reliable ground-truth labels for evaluating the automated methods.

\section{Results}

\subsection{Descriptive Statistics}
\begin{figure*}[t]  \centering
  \includegraphics[width=\textwidth]{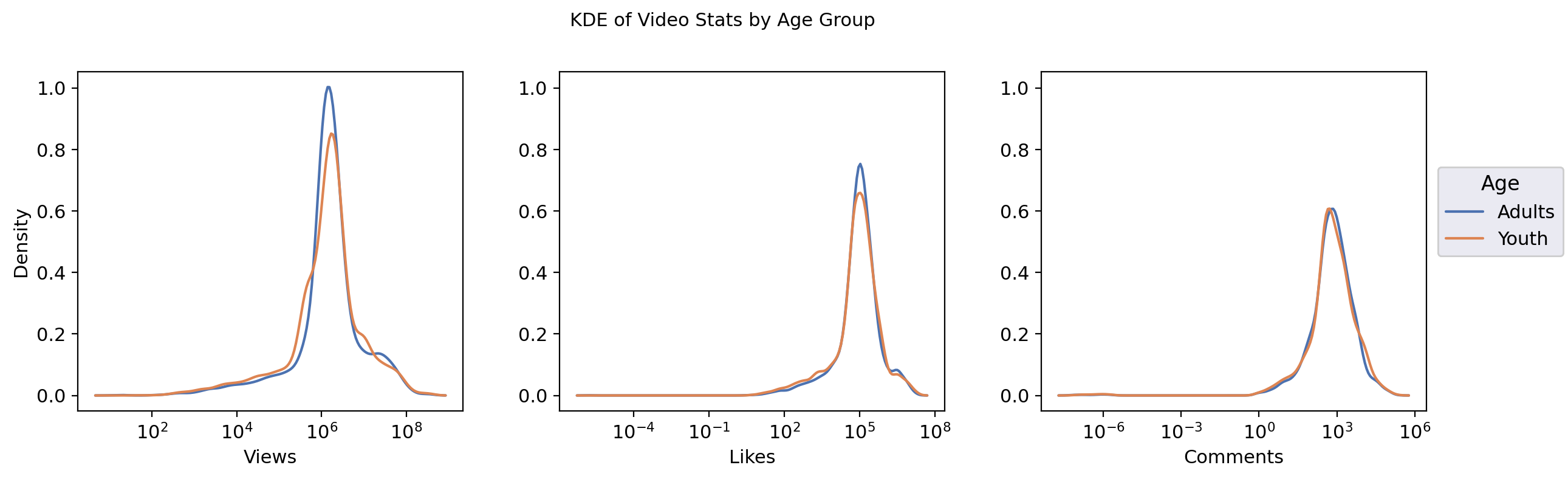}
  \caption{Distributions of metrics for videos collected by accounts belonging to the Adult and Youth group.}
  \label{fig:kde_views}
\end{figure*}

We analyzed the sample of videos collected with the two sets of accounts. As shown in Figure~\ref{fig:kde_views}, the distributions of views, likes, and comments are nearly identical across the two groups, suggesting that the two account modalities expose users to videos with similar engagement on the platform. We did not observe differences in the data collected on weekends or weekdays.



\subsection{Prevalence of toxic comments}
To analyze harmful content in the comments shown below videos, we compare the distribution of toxicity across all collected comments as well as the maximum toxicity of individual comments for each video.

\begin{figure}[!t]
  \centering
\includegraphics[width=\columnwidth]{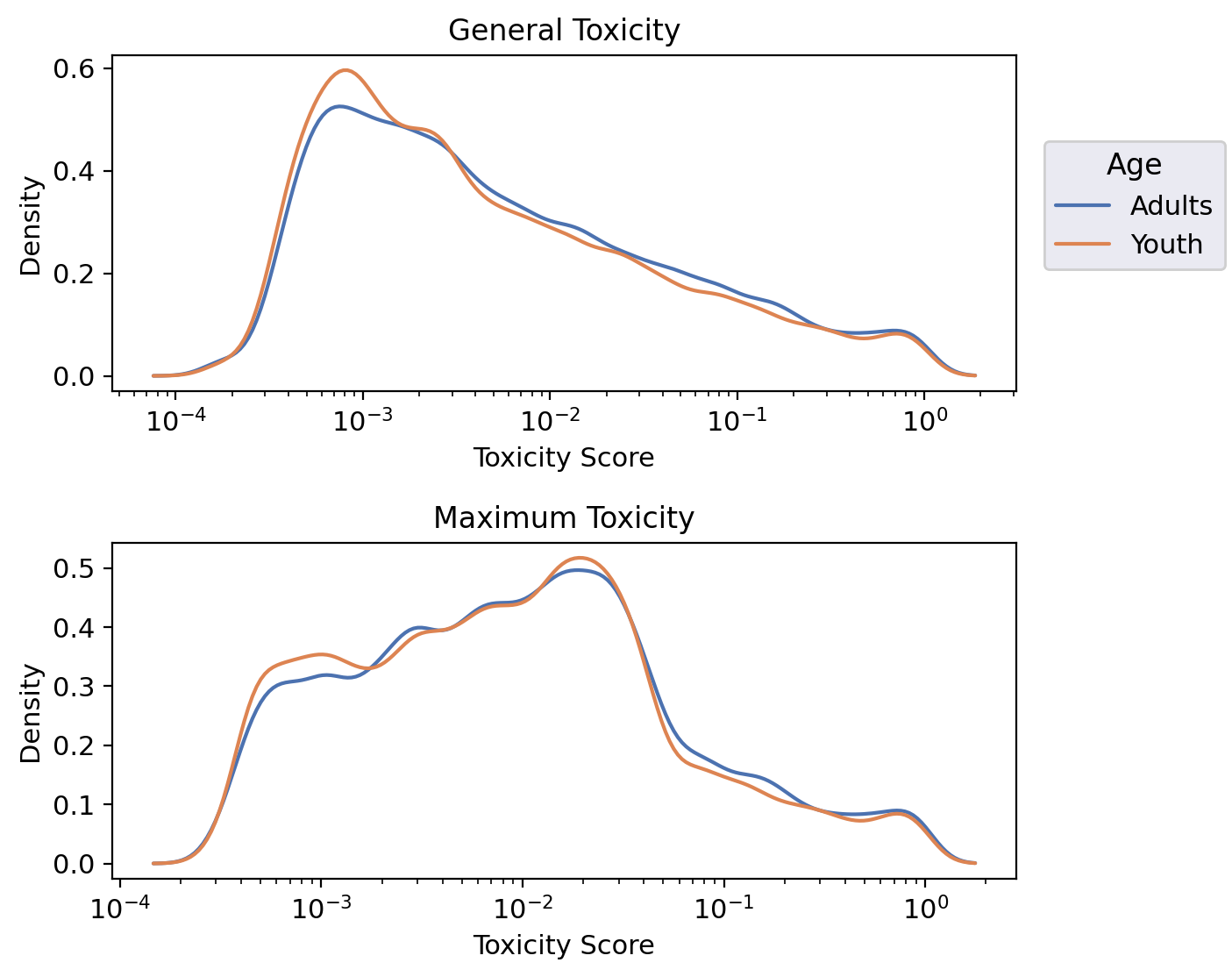}
  \caption{Distribution of toxicity for all comments (top) and the most toxic comment (bottom) below videos collected with the two groups of accounts. }
  \label{fig:avg_tox}
\end{figure}

As shown in Figure~\ref{fig:avg_tox}, the distributions of toxicity scores for both adult and youth accounts exhibit very similar patterns across the two types of analysis--overall toxicity of all comments and maximum toxicity per video. Assuming a commonly used threshold of 0.5–0.7~\cite{10.1145/3313831.3376548} to classify a comment as toxic, the vast majority of comments in both groups would not be considered harmful, as the 95th percentile of toxicity is below 0.3 (0.26-0.28) for videos in both groups of accounts. Median values are also very similar in the two samples (\~0.03), for both analyses. 

These small differences suggest comparable levels of exposure to toxic content in the comment sections of videos shown to both adults and youth.  

\subsection{Estimated prevalence of harmful videos}

\begin{figure}[!t]
  \centering
  \includegraphics[width=\columnwidth]{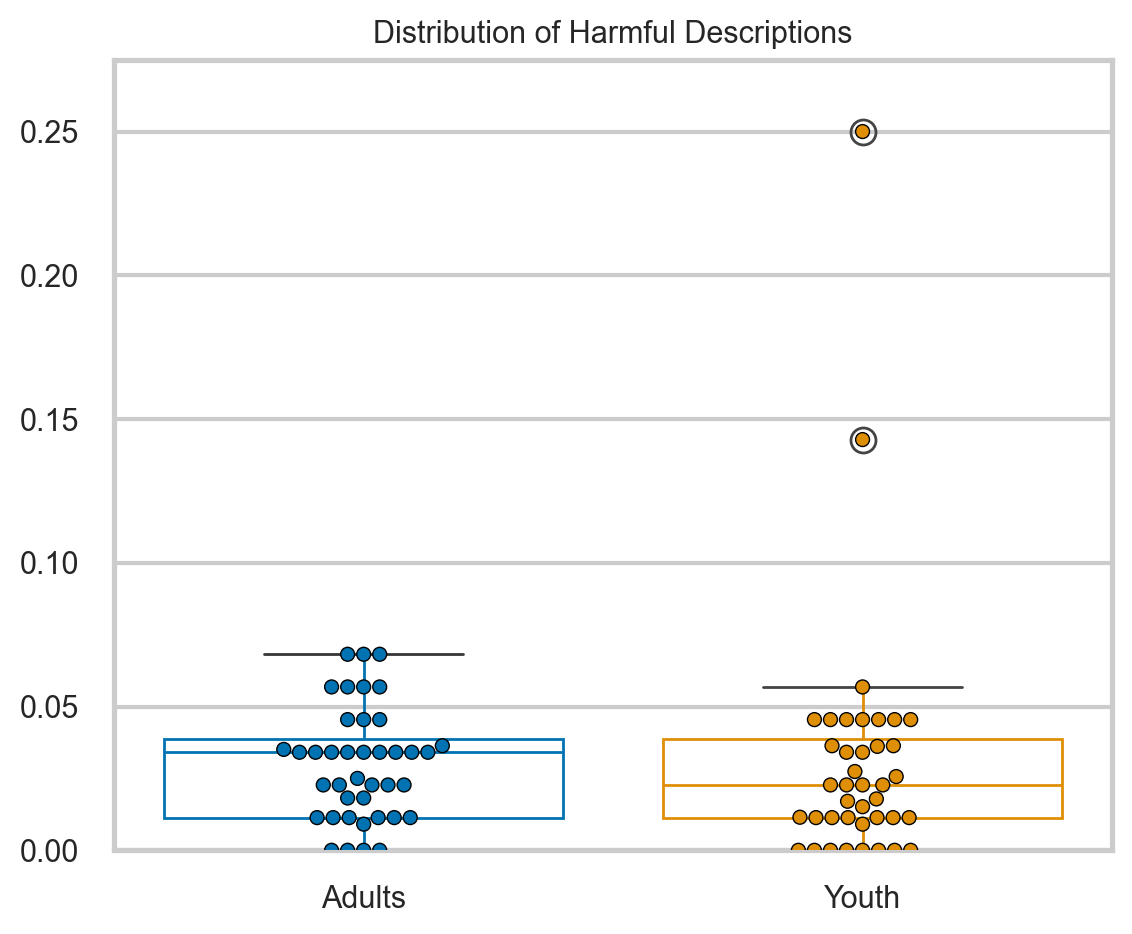}
  \caption{Distribution of estimated proportion of harmful videos shown to each account in the two groups, using labels provided by GPT-4o. Median values are 0.034 for Adults and 0.023 for Youths.}
  \label{fig:desc_prop}
\end{figure}

We estimated the prevalence of harmful videos shown to different accounts by using GPT-4o to annotate content based solely on video descriptions.
As shown in Figure~\ref{fig:desc_prop}, fewer than 10\% of the videos were predicted to be harmful for both adult and youth accounts during passive FYF scrolling. However, two youth accounts exhibited notably higher proportions, with about 14\% and 25\% of their videos labeled as harmful.

When focusing on active keyword-based searches using harm-related terms, the estimated prevalence of harmful content increases substantially for both groups. Specifically, 28.44\% of videos surfaced for adult accounts and 27.91\% for youth accounts were classified as harmful, showing rates significantly higher than during passive exposure.

These findings suggest that TikTok's safety mechanisms may be insufficient in protecting younger users from exposure to potentially harmful content, particularly when users actively search for related material.

\subsection{Evaluating the performance of harmful content classifiers}

We then evaluated the performance of GPT-4o and VideoLLaMA3 by manually annotating the random sample of 100 videos--50 from each age group, with 25 per group collected via keyword-based searches. The Fleiss' Kappa coefficient for inter-annotator agreement was 0.45, indicating moderate agreement. The raw mean percent agreement was 87.3\% \cite{landis1977measurement}. The gap between these measures reflects the high expected chance agreement in our binary harmful-content classification task.

Manual annotation identified genuinely harmful content in 41 of the 100 videos—26 in the Adult sample and 15 in the Youth sample. Against this manual labelling, GPT-4o achieved a precision of 59\%, and VideoLLaMA3 achieved 58\%.

\section{Discussion}






We analyzed a dataset of over 7,000 TikTok videos collected from age-specific sockpuppet accounts using both passive and active interaction protocols. Our findings indicate that adult and youth accounts were exposed to highly similar content, with fewer than 1 in 10 videos estimated to be harmful. This suggests that TikTok's moderation mechanisms may not meaningfully differentiate between age groups in practice. Despite this, automated detection of harmful content remains challenging: experiments with VLMs showed limited precision, with GPT-4o — operating on text only - outperforming VideoLLaMA3, which had access to video content. 

Our work presents promising results towards building a scalable and reproducible approach for auditing exposure to harmful content on recommender-driven platforms. By systematically comparing the content shown to adult and youth users, one can uncover tangible shortcomings in age-based content moderation, especially concerning the search function. Our results suggest that platforms like TikTok should invest greater effort in strengthening youth safety protections and ensuring greater transparency in how content is being moderated for these specific audiences.

Future research will expand the scope of this study by including additional countries to assess cross-cultural differences in content moderation. We also plan to scale up the experiment with a larger and more diverse set of accounts, including personalized profiles that better simulate real user behavior. To improve ecological validity, future data collection will incorporate the mobile interface, reflecting the primary mode of user interaction on TikTok. Finally, we aim to develop and integrate more accurate prediction models for detecting harmful content, leveraging multimodal signals more effectively.



\section*{Ethical Considerations and Limitations}
Our study involves the collection and analysis of publicly available content from TikTok through automated scraping techniques. This data collection was conducted solely for academic research purposes, with the aim of auditing systemic risks on a very large online platform, an activity explicitly permitted under the DSA. In line with the DSA's provisions, our work contributes to the broader public interest by examining the effectiveness of safety enforcement mechanisms, particularly as they relate to youth protection and the dissemination of harmful content.
We do not attempt to identify or track individual users, nor do we collect personal data or metadata that could be used to do so. Furthermore, none of the collected data is released or shared in a way that could compromise user anonymity or platform integrity. As our research does not involve human subjects or interactions with real users, it does not require approval from an institutional ethics review board.

Nevertheless, this work has several limitations. First, our classification of harmful content relies on a limited set of LLMs. Second, our analysis focuses exclusively on the Italian language and user experience within Italy, which may limit the generalizability of our findings to other linguistic or cultural environments. Additionally, we simulate user behavior using a relatively small number of sockpuppet accounts, which may not fully reflect the diversity of real user interactions on the platform. 


\bibliography{custom}

\appendix

\section{Details of Harmful Content Classification}
\label{sec:appendix}
\subsection{Framework}
Table~\ref{tab:categories} summarizes the harmful content categories used in our analysis, 
based on TikTok's official Community Guidelines.

\begin{table*}[!htbp]
  \centering
  \caption{Harmful content categories based on TikTok community guidelines}
  \label{tab:categories}
  \small
  \begin{tabular}{p{4.2cm}|p{9.8cm}}
    \hline
    \textbf{Category} & \textbf{Definition} \\
    \hline
    Disordered Eating and Body Image & Showing or promoting disordered eating and dangerous weight loss behaviors, or facilitating the trade or marketing of weight loss or muscle gain products. \\
    \hline
    Suicide and Self-Harm & Showing, promoting, or sharing plans for suicide or self-harm. \\
    \hline
    Dangerous Activity and Challenges & Showing or promoting dangerous activity and challenges. \\
    \hline
    Nudity and Body Exposure & Bare genitalia, buttocks, breasts of women and girls, sheer clothing, or semi-nudity or significant body exposure of young people. \\
    \hline
    Sexually Suggestive Content & Content by young people that intends to be sexually suggestive, including intimate kissing, sexualized framing, or behavior, and sexually explicit language. \\
    \hline
    Shocking and Graphic Content & Gory, gruesome, disturbing, or extremely violent content. \\
    \hline
    Sexual and Physical Abuse & Showing, promoting, or engaging in adult or youth sexual or physical abuse or exploitation. \\
    \hline
    Human Trafficking and Smuggling & Recruitment and transport of victims using force, fraud, or coercion, or smuggling someone across borders illegally for profit. \\
    \hline
    Hate Speech or Behavior & Hate speech, hateful behavior, or promotion of hateful ideologies. \\
    \hline
    Gambling & Betting money or items of value on uncertain outcomes, or engaging in gambling-like behaviors such as social casinos. \\
    \hline
    Alcohol, Tobacco, and Drugs & Trade, possession, or use of drugs, alcohol, or tobacco products, including regulated substances and youth use. \\
    \hline
    Harassment and Bullying & Violent threats, hate speech, sexual exploitation, doxxing, or other serious forms of harassment. \\
    \hline
    Integrity and Authenticity & Misinformation, election integrity issues, AI-generated deception, fake engagement, spam, and unoriginal or deceptive content. \\
    \hline
  \end{tabular}
\end{table*}

\subsection{GPT-4o Prompt}

  \begin{quote}
  Classify the following TikTok video \textbf{descriptions} as 'harmful' or 'not harmful' under this harmful content framework:
  \begin{itemize}
    \item Disordered Eating and Body Image
    \item Suicide and Self-Harm
    \item Dangerous Activity and Challenges
    \item Nudity and Body Exposure
    \item Sexually Suggestive Content
    \item Shocking and Graphic Content
    \item Hate Speech or Behavior
    \item Sexual and Physical Abuse
    \item Human Trafficking and Smuggling
    \item Gambling
    \item Alcohol, Tobacco, and Drugs
    \item Integrity and Authenticity
    \item Harassment and Bullying
  \end{itemize}

  Reply in the format: \texttt{harmful/not harmful} based on whether the TikTok falls under one of these groups.
  \end{quote}

\subsection{VideoLLaMA Prompt}
    
  \begin{quote}
  Classify the following TikTok videos as 'harmful' or 'not harmful' under this harmful content framework:
  \begin{itemize}
    \item Disordered Eating and Body Image
    \item Suicide and Self-Harm
    \item Dangerous Activity and Challenges
    \item Nudity and Body Exposure
    \item Sexually Suggestive Content
    \item Shocking and Graphic Content
    \item Hate Speech or Behavior
    \item Sexual and Physical Abuse
    \item Human Trafficking and Smuggling
    \item Gambling
    \item Alcohol, Tobacco, and Drugs
    \item Integrity and Authenticity
    \item Harassment and Bullying
  \end{itemize}

  Reply in the format: \texttt{harmful/not harmful} based on whether the TikTok falls under one of these groups.
  \end{quote}

\section{Keywords for Active Searching}
\label{sec:appendix}

Table~\ref{tab:keywords} lists the Italian keywords associated with each harmful content category, derived from TikTok's Youth Safety and Well-Being Guidelines. 
These keywords were used in the active searching experiments to retrieve videos potentially related to sensitive topics. 

\begin{table*}[!t]
  \centering
  \caption{Keyword examples associated with TikTok harmful content categories (in Italian)}
  \label{tab:keywords}
  \small
  \begin{tabular}{p{5cm}|p{3cm}|p{3cm}|p{3cm}}
    \hline
    \textbf{Category} & \textbf{Term 1} & \textbf{Term 2} & \textbf{Term 3} \\
    \hline
    Disturbi legati all'alimentazione e immagine corporea & Diete restrittive & perdere peso & chirurgia estetica \\
    \hline
    Challenge e attività pericolose & challenge pericolose & attività pericolose & tendenza \\
    \hline
    Nudità ed esposizione corporea & nudità & esposizione corporea & biancheria intima \\
    \hline
    Contenuti sessualmente allusivi & baci intimi & sesso & prodotti per il sesso \\
    \hline
    Contenuti grafici e scioccanti & sangue & combattimenti & disastro naturale \\
    \hline
    Gioco & gioco & casinò & scommesse \\
    \hline
    Alcol e tabacco e droghe & droghe & tabacco & alcol \\
    \hline
  \end{tabular}
\end{table*}

\end{document}